\documentclass[preprint,showkeys,showpacs]{revtex4-1}
\usepackage{amssymb,amsmath,graphicx}
\renewcommand{\figurename}{Fig.}

\begin{document} 
\title{Properties of the diffusion and drift kinetic coefficients in momentum space for a cold Fermi system}
\author{Sergiy V. Lukyanov}
\affiliation{Institute for Nuclear Research, 03680 Kyiv, Ukraine}
\pacs{21.60.-n, 21.60.Ev, 24.30.Cz}
\date{\today}

\begin{abstract}
Using the methods of kinetic theory expressions for the diffusion and drift coefficients for a cold Fermi system are obtained.
Their dependences on the momentum are calculated for the step distribution function as well as in the case of excitation of a particle-hole pair.
\end{abstract}

\keywords{kinetic theory; Fermi system; diffusion approach; diffusion and drift coefficients.}

\maketitle

\section{Introduction}

Recently, in a number of papers \cite{BaWo.AP.2019,KoLu.UPJ.2014,KoLu.IJMP.2015,Lu.IJMP.2021} the processes of relaxation of collective excitations 
in a Fermi system were considered in the framework of the diffuse approximation. This approximation is based on a nonlinear diffusion equation in 
momentum space which contains the kinetic coefficients of diffusion and drift. In the general case the diffusion and drift coefficients depend not 
only on time but also on momentum. However, an approximation of constant kinetic coefficients is often used. In this case the nonlinear diffusion 
equation can be simplified and solved exactly \cite{Wo.PRL.1982}. This solution allows us to study the evolution of the Fermi system in time and 
obtain an expression for the relaxation time of both collective excitations and excitations of the particle-hole type in atomic nuclei 
\cite{BaWo.AP.2019,Wo.PRL.1982}. 

The approximation of constant kinetic coefficients is rather rough since it does not take into account their rather significant dependence on momentum
\cite{KoLu.UPJ.2014,KoLu.IJMP.2015}. The study of the apparent momentum dependence of the kinetic coefficients is actually a separate problem.
In this case, when it is considered by the methods of the kinetic theory it is necessary to solve the problem with the ninefold collision integral by means of which the coefficients are determined. To simplify calculations at low temperatures the small momentum transfer approximation is used when particles are scattered near the Fermi surface.  However, this is not enough to obtain a physically correct result. It is necessary to make an assumption about the nature 
of the interaction of the scattering particles. Previously, the isotropic probability approximation was considered a good approximation
nucleon scattering \cite{AbKh.UFN.1958,KoLuPlSh.PRC.1998}. However, in this case when calculating the kinetic coefficients divergent integral expressions 
are obtained. In previous papers, we showed that this can be avoided by imposing the short range conditions on the internucleon potential \cite{KoLu.IJMP.2015}. In particular for the Gauss potential it was possible to obtain convergent expressions for the kinetic coefficients
and calculate their numerical values and temperature dependences based on such phenomenological parameters of internucleon interaction as the potential depth
and its effective radius. At the same time, the calculation of the kinetic coefficients for zero temperature was not carried out due to the arising technical difficulties.

This work is devoted to solving these technical difficulties and to the exact calculation of the kinetic coefficients of diffusion and drift at zero temperature of the Fermi system and in the case of excitation of a particle-hole pair.

\section{Kinetic diffusion and drift coefficients}

Consideration will begin using the results already obtained in our previous papers. 
Let us write down the general expressions for the kinetic coefficients of diffusion and drift \cite{LiPi.bp2.1980,KoLu.IJMP.2015}
\begin{equation}
D_p(\mathbf{p})=\frac{1}{6}\int\frac{gd\mathbf{s}}{(2\pi \hbar )^{3}}\ s^{2}\ W(\mathbf{p},\mathbf{s}),  
\label{dpdef}
\end{equation}
\begin{equation}
K_p(\mathbf{p})=\frac{m}{p}\left(\nabla_p D_p(\mathbf{p})- A(\mathbf{p})\right),
\label{kpdef}
\end{equation}
where the corresponding integral expressions are \cite{KoLu.IJMP.2015}
\begin{equation}
A(\mathbf{p})=\int\frac{gd\mathbf{s}}{(2\pi \hbar )^{3}}\ \hat{\mathbf{p}} \mathbf{s}\ 
W(\mathbf{p},\mathbf{s}),
\label{adef}
\end{equation}
\begin{equation}
W(\mathbf{p},\mathbf{s})\approx \frac{2g}{m^{2}}\frac{d\sigma }{d\Omega}(\mathbf{s}^{2}) 
\int d\mathbf{p}_{2}\ d\mathbf{p}_{4}\ \tilde{f}(\mathbf{p}_{2})
f(\mathbf{p}_{4})\delta \left( \mathbf{p}_{2}-\mathbf{p}_{4}-\mathbf{s}\right)\delta\left(\epsilon_{2}-\epsilon_{4} 
-\frac{\mathbf{p\cdot }\mathbf{s}}{m}\right).
\label{gaindef}
\end{equation}
Here $g$ is the spin-isospin degeneracy factor, $m$ is the nucleon mass, $d\sigma/d\Omega$ is the nucleon scattering cross section,
$\epsilon_j=p_j^2/2m$ is the kinetic energy. The blocking factor that takes into account the Pauli principle is indicated by a tilde over the distribution function $\tilde{f}(\mathbf{p})=1-f(\mathbf{p})$.

Recall that in the microscopic calculation of the kinetic coefficients of diffusion $D_p(\mathbf{p})$ and drift $K_p(\mathbf{p})$ one considers 
the scattering of nucleons with initial momenta $\mathbf{p}_1$ and $\mathbf{p}_2$ into the final states $\mathbf{p}_3$ and $\mathbf{p}_4$. 
Nucleon scattering occurs near the Fermi surface, so the cross section $d\sigma/d\Omega$ depends on the square of the small momentum transferred
$\mathbf{s}=\mathbf{p}_1-\mathbf{p}_3$ \cite{KoLu.IJMP.2015,AbKh.UFN.1958}. 

After substituting Eq. (\ref{gaindef}) into the Eqs. (\ref{dpdef}) and (\ref{adef}) and integrating over the small momentum transfer $\mathbf{s}$, 
we get
\begin{eqnarray}
D_p(\mathbf{p}) \approx \frac{2g^2}{6m^{2}(2\pi\hbar)^{3}} \int d\mathbf{p}_{2}d\mathbf{p}_{4}\ (\mathbf{p}_{2}-\mathbf{p}_{4})^2\
\frac{d\sigma }{d\Omega } \left((\mathbf{p}_{2}-\mathbf{p}_{4})^2\right) \tilde{f}(\mathbf{p}_{2})f(\mathbf{p}_{4}) 
\nonumber \\
\times \delta \left(\epsilon _{2}-\epsilon _{4} 
-\frac{\mathbf{p}}{m}(\mathbf{p}_{2}-\mathbf{p}_{4})\right),  
\label{dp2}
\end{eqnarray}
\begin{eqnarray}
A(\mathbf{p}) \approx \frac{2g^2}{m^{2}(2\pi\hbar)^{3}} \int d\mathbf{p}_{2}d\mathbf{p}_{4}\ \hat{\mathbf{p}} (\mathbf{p}_{2}-\mathbf{p}_{4})\ 
\frac{d\sigma}{d\Omega}\left((\mathbf{p}_{2}-\mathbf{p}_{4})^2\right)\tilde{f}(\mathbf{p}_{2})f(\mathbf{p}_{4}) 
\nonumber \\
\times\delta\left(\epsilon_{2}-\epsilon_{4}-\frac{\mathbf{p}}{m}(\mathbf{p}_{2}-\mathbf{p}_{4})\right).  
\label{ap2}
\end{eqnarray}

Carrying out the transformations described in detail in Appendix A, we obtain general expressions for the diffusion coefficient $D_p(p)$ and the integral 
value $A(p)$, which is included in the definition of the drift coefficient Eq. (\ref{kpdef}), for the case of a spherically symmetric distributions 
$f(\mathbf{p})=f(p)$ of nucleons in the momentum space
\begin{eqnarray}
D_p(p)&\approx&\frac{2g^2}{3m(2\pi\hbar)^{3}} 
\int_{0}^{\infty} k^5 dk\ \int d\Omega_{k}\ f\left(\sqrt{k^2+p^2+2kp\cos\theta_k}\right)
\nonumber \\
&& \times \int d\Omega_{q} 
\left\{\left(1-\cos\widehat{\mathbf{q}\mathbf{k}}\right) 
\frac{d\sigma}{d\Omega}\left(2k^2(1-\cos \widehat{\mathbf{q}\mathbf{k}})\right)\right.
\tilde{f}\left(\sqrt{k^2+p^2+2kp\cos\theta_q}\right)
\nonumber \\
&& + \left(1+\cos\widehat{\mathbf{q}\mathbf{k}}\right) 
\frac{d\sigma}{d\Omega}\left(2k^2(1+\cos\widehat{\mathbf{q}\mathbf{k}})\right)
\left.\tilde{f}\left(\sqrt{k^{2}+p^{2}-2kp\cos\theta_q}\right)\right\},
\label{dp6-a}
\end{eqnarray}
\begin{eqnarray}
A(p)&\approx&\frac{2g^2}{m(2\pi\hbar)^{3}} 
\int_{0}^{\infty} k^4 dk\ \int d\Omega_{k}\ f\left(\sqrt{k^2+p^2+2kp\cos\theta_k}\right)
\nonumber \\
&& \times \int d\Omega _{q}\ \left\{\left(\cos\theta_q-\cos\theta_k\right) 
\frac{d\sigma}{d\Omega}\left(2k^2(1-\cos \widehat{\mathbf{q}\mathbf{k}})\right)\right.
\tilde{f}\left(\sqrt{k^2+p^2+2kp\cos\widehat{\mathbf{q}\mathbf{p}}}\right)
\nonumber \\
&& -\left(\cos\theta_q+\cos\theta_k\right) 
\frac{d\sigma}{d\Omega}\left(2k^2(1+\cos \widehat{\mathbf{q}\mathbf{k}})\right)
\left.\tilde{f}\left(\sqrt{k^{2}+p^{2}-2kp\cos\widehat{\mathbf{q}\mathbf{p}}}\right) 
\right\}.
\label{a6-a}
\end{eqnarray}

Note, that in order for the integral expressions of the Eqs. (\ref{dp6-a}) and (\ref{a6-a}) to have limited values the following condition 
for the differential cross section must be satisfied:
\begin{equation}
\lim_{k\rightarrow\infty}\ k^5 \frac{d\sigma}{d\Omega}\left(2k^2(1 \pm\cos \widehat{\mathbf{q}\mathbf{k}})\right) =0.
\end{equation}
This condition is satisfied by the finite-radius inter-particle interaction with the following Gaussian form-factor 
$v(r)=v_{0}\exp (-r^{2}/2r_{0}^{2})$ which is appropriate for calculations of the in-medium cross-section within the
transport approaches.
The differential cross section $d\sigma /d\Omega $ in the first Born approximation is then given by \cite{Dav.b.65}
\begin{equation}
\frac{d\sigma (\mathbf{s}^{2})}{d\Omega}=
\frac{\pi m^{2}r_{0}^{6}v_{0}^{2}}{2\hbar ^{4}}
\exp \left(-4\mathbf{s}^{2}r_{0}^{2}/\hbar ^{2}\right),  
\label{pot}
\end{equation}
were $r_{0}$ and $v_{0}$ are the free parameters.

Substituting Eq. (\ref{pot}) into the integral expressions (\ref{dp6-a}) and (\ref{a6-a}) and performing transformations
which are described in detail in Appendix B, we obtain the final expressions for the diffusion coefficient $D_p(p)$ and
integral expression $A(p)$:
\begin{eqnarray}
D_p(p)&\approx &\frac{g^2 m r_{0}^{6}v_{0}^{2}}{3\hbar^7}
\int_{0}^{\infty} k^5 dk \int_{-1}^1 dx \int_{-1}^1 dy\ \exp(-\alpha(k,x,y)) \tilde{f}(q_{x}) f(q_{y})
\nonumber \\
&&\times 
\left[(1-xy)I_0(\beta(k,x,y))-\sqrt{(1-x^2)(1-y^2)}I_1(\beta(k,x,y))\right],
\label{dp13-b} \\
A(p)&\approx &\frac{g^2 m r_{0}^{6} v_{0}^{2}}{\hbar^7}
\int_{0}^{\infty} k^4 dk \int_{-1}^1 dx \int_{-1}^1 dy\ \exp(-\alpha(k,x,y)) 
\nonumber \\
&&\times \tilde{f}(q_{x}) f(q_{y}) (x-y) I_0\left(\beta(k,x,y)\right),
\label{ap13-b}
\end{eqnarray}
where the functions of three variables $\alpha(k,x,y)$ and $\beta(k,x,y)$ have the form
\begin{equation}
\alpha(k,x,y)=8k^2r_0^2(1-xy)/\hbar^2, 
\quad
\beta(k,x,y)=8k^2r_0^2\sqrt{(1-x^2)(1-y^2)}/\hbar^2.
\end{equation}
Also here are the expressions
\begin{equation}
q_{x}=\sqrt{p^2+k^2+2pkx}, \quad q_{y}=\sqrt{p^2+k^2+2pky}.
\end{equation}
$I_n(x)$ are modified Bessel functions of the first kind.

\section{Kinetic coefficients for a cold Fermi system}

Consider the Fermi system at zero temperature ($T=0$), which is described by the step distribution function in the momentum space
\begin{equation}
f(p)=\theta(p_F^2-p^2).
\label{fus}
\end{equation}

For the step distribution function (\ref{fus}), the integration contribution is limited to a sphere with radius $p_F$,
therefore, when integrating in the expressions (\ref{dp13-b}) and (\ref{ap13-b}), the product $\tilde{f}(q_{x}) f(q_{y})$
highlights a certain area, which can be taken into account by means of certain minimum and maximum values of the integration limits.

The expressions for the diffusion coefficient $D_p(p)$ and the quantity $A(p)$ can be represented in the form convenient for further consideration
\begin{equation}
D_p(p)\approx\mathrm{const}\int_0^\infty dk\ k^5\ d(p,k),
\label{dpt0}
\end{equation}
\begin{equation}
A(p)\approx\mathrm{const}\ 3\int_0^\infty dk\ k^4\ a(p,k),
\label{apt0}
\end{equation}
where
\begin{equation}
\mathrm{const}=\frac{g^{2}mr_{0}^{6}v_{0}^{2}}{3 \hbar^{7}},
\end{equation}
\begin{equation}
d(p,k)= \int_{x_{\min}}^1 dx \int_{-1}^{y_{\max}}dy\  G(k,x,y),
\label{dpk}
\end{equation}
\begin{equation}
a(p,k)= \int_{x_{\min}}^1 dx\ \int_{-1}^{y_{\max}}dy\ F(k,x,y).
\label{apk}
\end{equation}
The minimum and maximum integration limits in the expressions (\ref{dpk}), (\ref{apk}) take the form
\begin{eqnarray}
x_{\min}=\min(1,\max(-1,z)), 
\quad
y_{\max}=\max(-1,\min(1,z)), 
\label{defxyminmax}
\end{eqnarray}
where the notation is defined as
\begin{equation}
z=\frac{p_F^2-p^2-k^2}{2pk}.
\label{defz}
\end{equation}
Accordingly, the integrand functions are written as
\begin{eqnarray}
G(k,x,y)&=&\exp(-\alpha(k,x,y)) 
\nonumber \\
&\times&\left[(1-xy)I_0(\beta(k,x,y))-\sqrt{(1-x^2)(1-y^2)}I_1(\beta(k,x,y))\right],
\label{gkxy} \\
F(k,x,y)&=&(x-y) \exp(-\alpha(k,x,y)) I_0(\beta(k,x,y)),
\label{fkxy}
\end{eqnarray}

The expressions (\ref{dpt0}) and (\ref{apt0}) at zero momentum are zero $D_p(0)=0$, $A(0)=0$. 
As the momentum increases to infinitely large values both quantities asymptotically approach zero values
\begin{equation}\label{lim-da}
\lim_{p\rightarrow\infty}D_p(p)=0, \qquad \lim_{p\rightarrow\infty}A(p)=0.
\end{equation}

\begin{figure}
\begin{center}
\includegraphics[scale=0.5,clip]{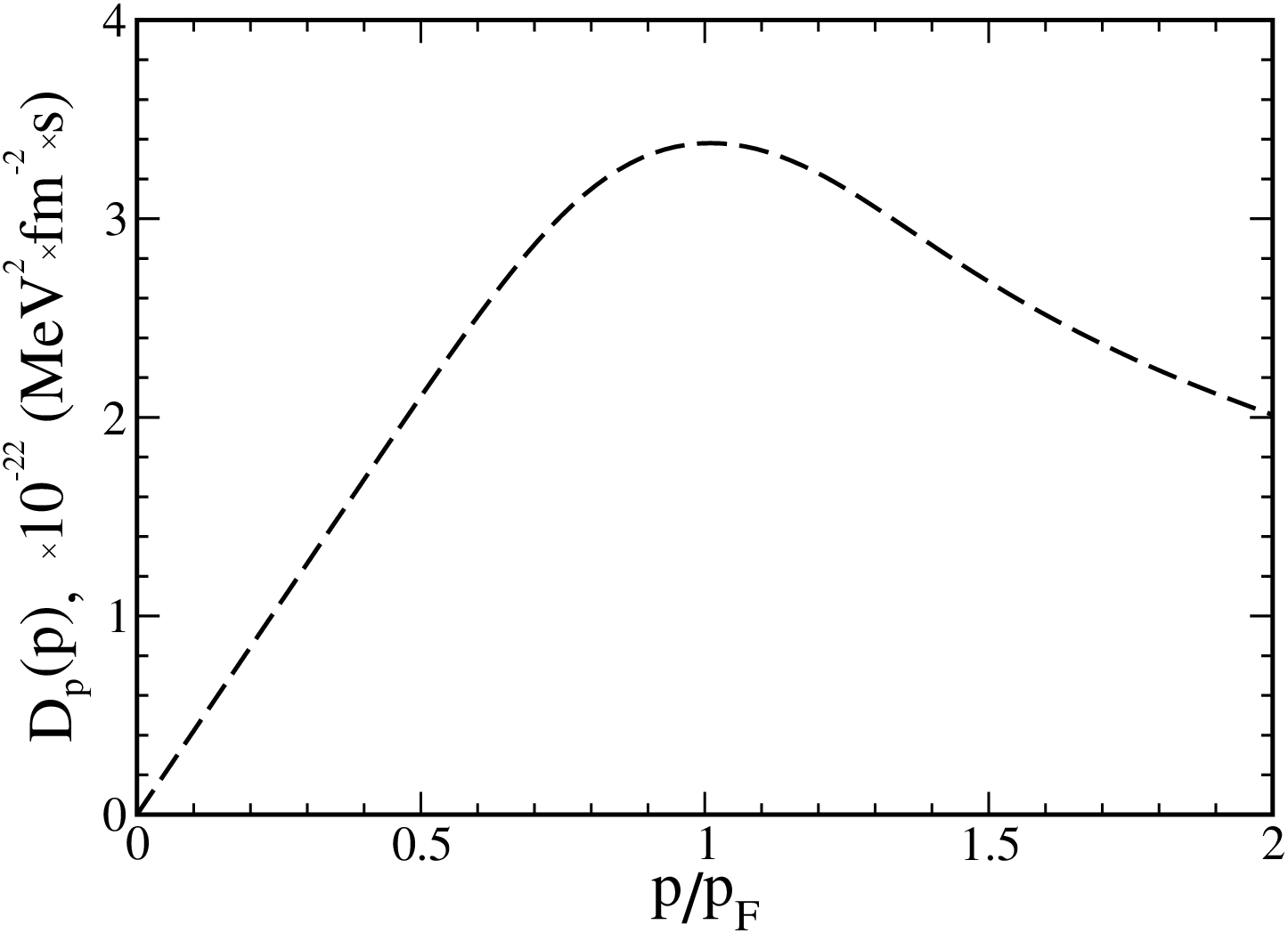}
\caption{Dependence of the diffusion coefficient $D_p(p)$ on the relative momentum $p/p_F$ for a cold Fermi system
described by the step distribution function (\ref{fus}).}
\label{fig1}
\end{center}
\end{figure}

\figurename~\ref{fig1} shows the diffusion coefficient $D_p(p)$ versus relative momentum (in units of the Fermi momentum $p_F$)
for the cold Fermi system ($T=0$), which are obtained in accordance with the expressions (\ref{dpt0})-(\ref{fkxy}).
For clarity, all the results were obtained for a light nucleus with a mass number $A=16$, because in the region of medium and heavy atomic nuclei 
the calculations are not representative. The Fermi energy was chosen to be typical for nuclear matter $\epsilon_F=37$ MeV.
As you can see, the diffusion coefficient is non-zero and has a maximum, which is located in the region of the Fermi surface and is approximately
$D_{p,0}(p_F)\approx 3.38\cdot 10^{-22}$ MeV$^{2}\cdot$fm$^{-2}\cdot$s. 
\begin{figure}
\begin{center}
\includegraphics[scale=0.5,clip]{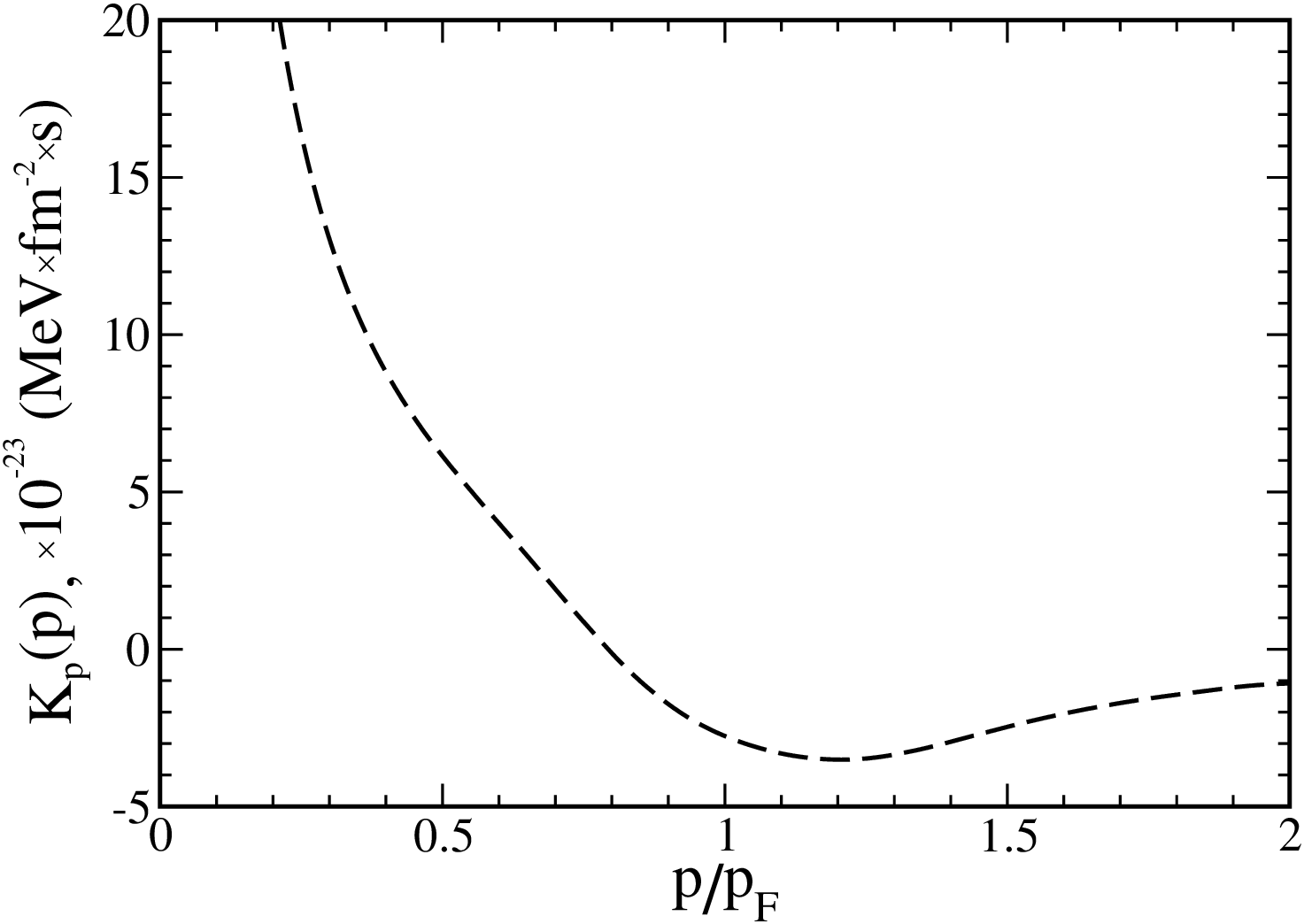}
\caption{Dependence of the drift coefficient $K_p(p)$ on the relative momentum $p/p_F$ for a cold Fermi system described 
by the step distribution function (\ref{fus}).}
\label{fig2}
\end{center}
\end{figure}
The drift coefficient in \figurename~\ref{fig2} is also different from zero and has a minimum with a negative sign which is localized near the Fermi surface 
($p/p_F\approx 1.2$). On the Fermi surface the value of the drift coefficient is
$K_{p,0}(p_F)\approx -2.76\cdot 10^{-23}$ $\text{MeV}\cdot\text{fm}^{-2}\cdot\text{s}$. In accordance with the expression (\ref{kpdef}) 
the coefficient drift tends to an infinitely large positive value as the momentum decreases to zero. Asymptotic dependences of the diffusion 
$D_p(p)$ and drift $K_p(p)$ coefficients with an infinite increase in momentum are consistent with their limiting zero values (\ref{lim-da}).

It is interesting to compare the values of $D_{p,0}(p_F)$ and $K_{p,0}(p_F)$ with the corresponding values of the diffusion and drift 
coefficients given in \cite{Wo.PRL.1982}: $D=20\times 10^{23}$ $\text{MeV}^2\cdot\text{s}^{-1}$, $v=-5\times 10^{23}$ $\text{MeV}^2\cdot\text{s}^{-1}$.
It should be noted here that our approach considers the kinetic coefficients for the diffusion equation in partial derivatives with respect 
to time and momentum \cite{KoLu.UPJ.2014,KoLu.IJMP.2015}. This fact is denoted by the subscript "p" of the coefficients $D_p(p)$ and $K_p(p)$. 
In \cite{Wo.PRL.1982} the diffusion equation is written in partial derivatives with respect to time and energy. Therefore, the corresponding 
kinetic coefficients $D$ and $v$ have different values and dimensions. Comparing the diffusion equation for both cases we find
$$
D\approx D_{p,0}(p_F)\left(\frac{p_F}{m}\right)^2, \qquad v\approx K_{p,0}(p_F)\left(\frac{p_F}{m}\right)^2.
$$
Taking into account the value of the Fermi energy we obtain the value of the multiplication factor $p_F^2/m^2\approx 0.71\times 10^{46}$ 
$\text{fm}^{2}\cdot\text{s}^{-2}$. Substituting the obtained numerical values we get $D\approx 24\times 10^{23}$ $\text{MeV}^2\cdot\text{s}^{-1}$, 
$v\approx -2\times 10^{23}$ $\text{MeV}\cdot\text{s}^{-1}$. As we can see, the diffusion coefficient obtained in our approach is quite close to the 
value from \cite{Wo.PRL.1982}, while the drift coefficient differs by a factor of two and a half.

In accordance with the diffusion equation the difference from zero of the diffusion and drift coefficients means that even in an absolutely cold 
Fermi system momentum transfer processes and consequently a change of state are possible. The most rapid smearing of the step distribution will occur 
in the region of the Fermi surface and as you move away from it the smearing rates will decrease.

\section{Particle-hole excitation in a cold Fermi system}

Let us now consider the excitation of a particle-hole pair in a cold Fermi system. 
Such a state is described by the initial distribution function in the form
\begin{equation}
f_{\mathrm{in}}(p)=\left[\theta (p_{1}^{\prime 2}-p^2)+\theta(p^2-p_{2}^{\prime 2})\right] 
\theta(p_{F}^2-p^2)+\theta(p_{2}^2-p^2)\theta(p^2-p_{1}^2).
\label{fin}
\end{equation}
The distribution $f_{\mathrm{in}}(p)$ of Eq. (\ref{fin}) means the particle
located at $p_{1}<p<p_{2}$ and the hole excitation at $p_{1}^{\prime
}<p<p_{2}^{\prime }$ for fixed $p_{1}>p_{F}$ and $p_{2}^{\prime }<p_{F}$,
respectively. The intervals $\Delta p^{\prime }=p_{2}^{\prime
}-p_{1}^{\prime }$ and $\Delta p=p_{2}-p_{1}$ are derived from the
conditions
\begin{eqnarray}
\int_{0}^{p_{F}} \frac{4\pi g\mathcal{V}dp}{(2\pi \hbar )^{3}} 
p^{2}f_{\mathrm{in}}(p,t=0) &=&A-1,  
\nonumber \\
\int_{p_{F}}^{\infty} \frac{4\pi g\mathcal{V}dp}{(2\pi\hbar)^{3}} 
p^{2}f_{\mathrm{in}}(p,t=0) &=&1.  
\label{fin2}
\end{eqnarray}

For such a distribution function, the product $\tilde{f}(q_{x}) f(q_{y})$ has a slightly more complex character the general expression of which is given in Appendix B. This expression contains 12 terms that impose conditions on the integration limits $x_{\min,i}, x_{\max,i}$ and $y_{\min,i}, y_{\max,i}$, 
where $ i=1, 2, ..., $12. Expressions for these integration limits are given in Appendix C.
So, taking into account the above, we present expressions for the diffusion coefficient and the integral value
$A(p)$:
\begin{equation}
D_p(p)\approx\mathrm{const} \int_0^\infty dk\ k^5\ \left(\sum_{i=1}^{3} d_i(p,k) - \sum_{i=4}^{12} d_i(p,k)\right),
\label{dpint0}
\end{equation}
where
\begin{equation}
d_i(p,k)=\int_{x_{\min,i}}^{x_{\max,i}} dx\ \int_{y_{\min,i}}^{y_{\max,i}} dy\ G(k,x,y).
\label{dpik}
\end{equation}

\begin{equation}
A(p)\approx\mathrm{const}\ 3\int_0^\infty dk\ k^4\ \left(\sum_{i=1}^{3} a_i(p,k)
- \sum_{i=4}^{12} a_i(p,k)\right),
\label{apint0}
\end{equation}
where
\begin{equation}
a_i(p,k)=\int_{x_{\min,i}}^{x_{\max,i}} dx\ \int_{y_{\min,i}}^{y_{\max,i}} dy\ F(k,x,y).
\label{apik}
\end{equation}

\begin{figure}
\begin{center}
\includegraphics[scale=0.5,clip]{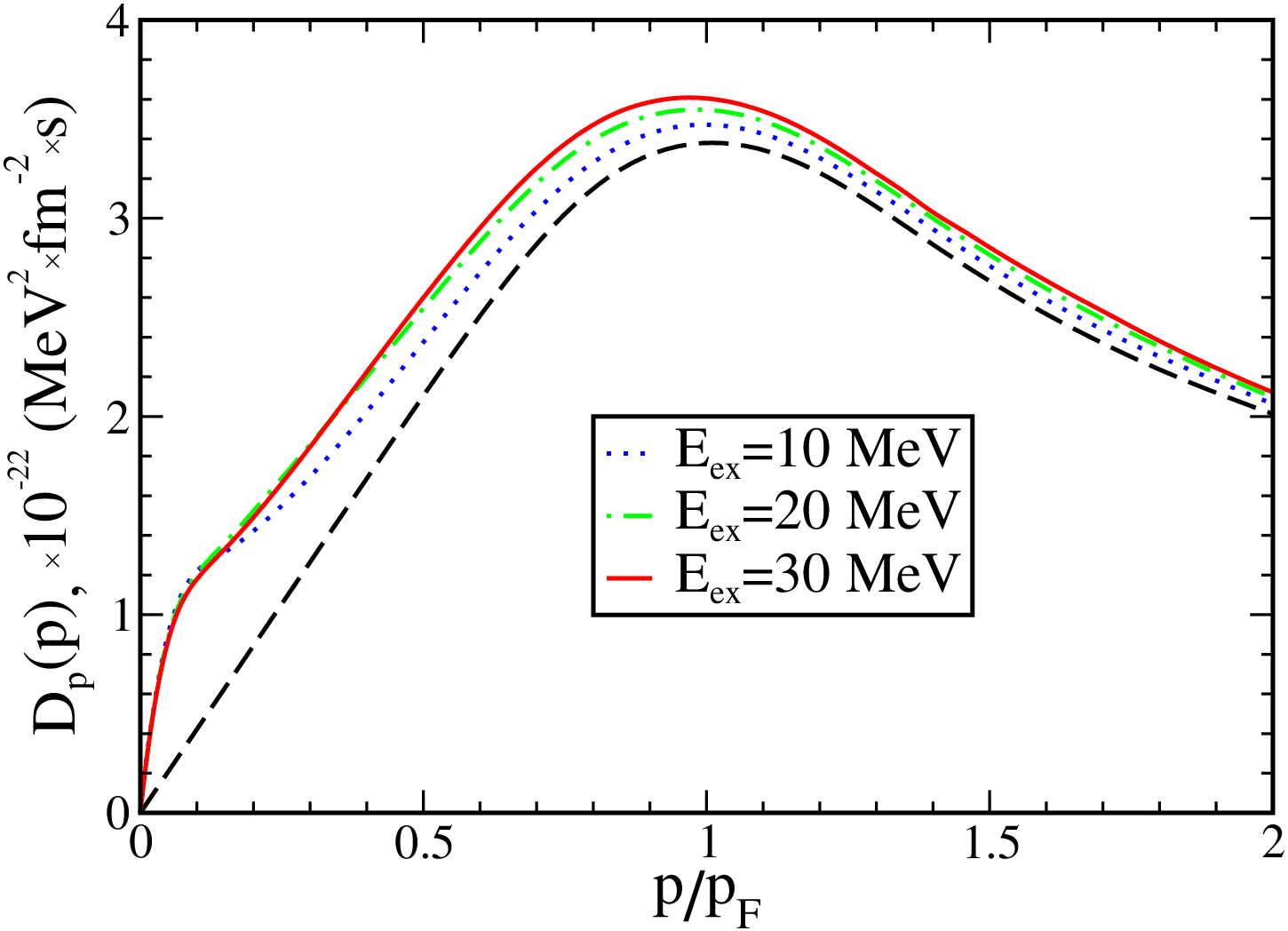}
\caption{Dependence of the diffusion coefficient $D_p(p)$ on the relative momentum $p/p_F$ in the case of excitation of a particle-hole pair 
for a Fermi system which is described by the distribution function (\ref{fus}) for case (a).}
\label{fig3}
\end{center}
\end{figure}

\begin{figure}
\begin{center}
\includegraphics[scale=0.5,clip]{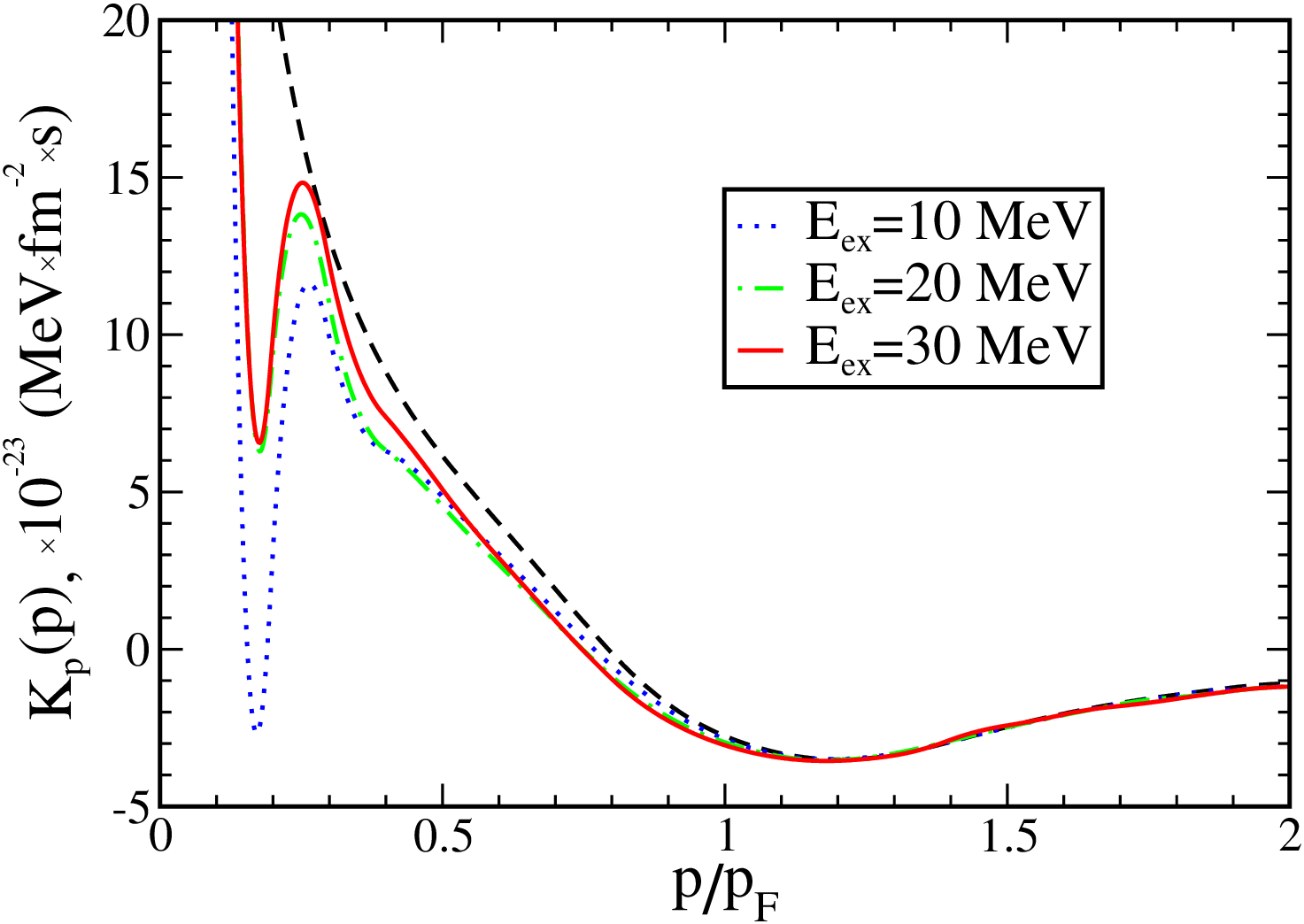}
\caption{Dependence of the drift coefficient $K_p(p)$ on the relative momentum $p/p_F$ in the case of excitation of a particle-hole pair 
for a Fermi system which is described by the distribution function (\ref{fus}) for case (a).}
\label{fig4}
\end{center}
\end{figure}

\begin{figure}
\begin{center}
\includegraphics[scale=0.5,clip]{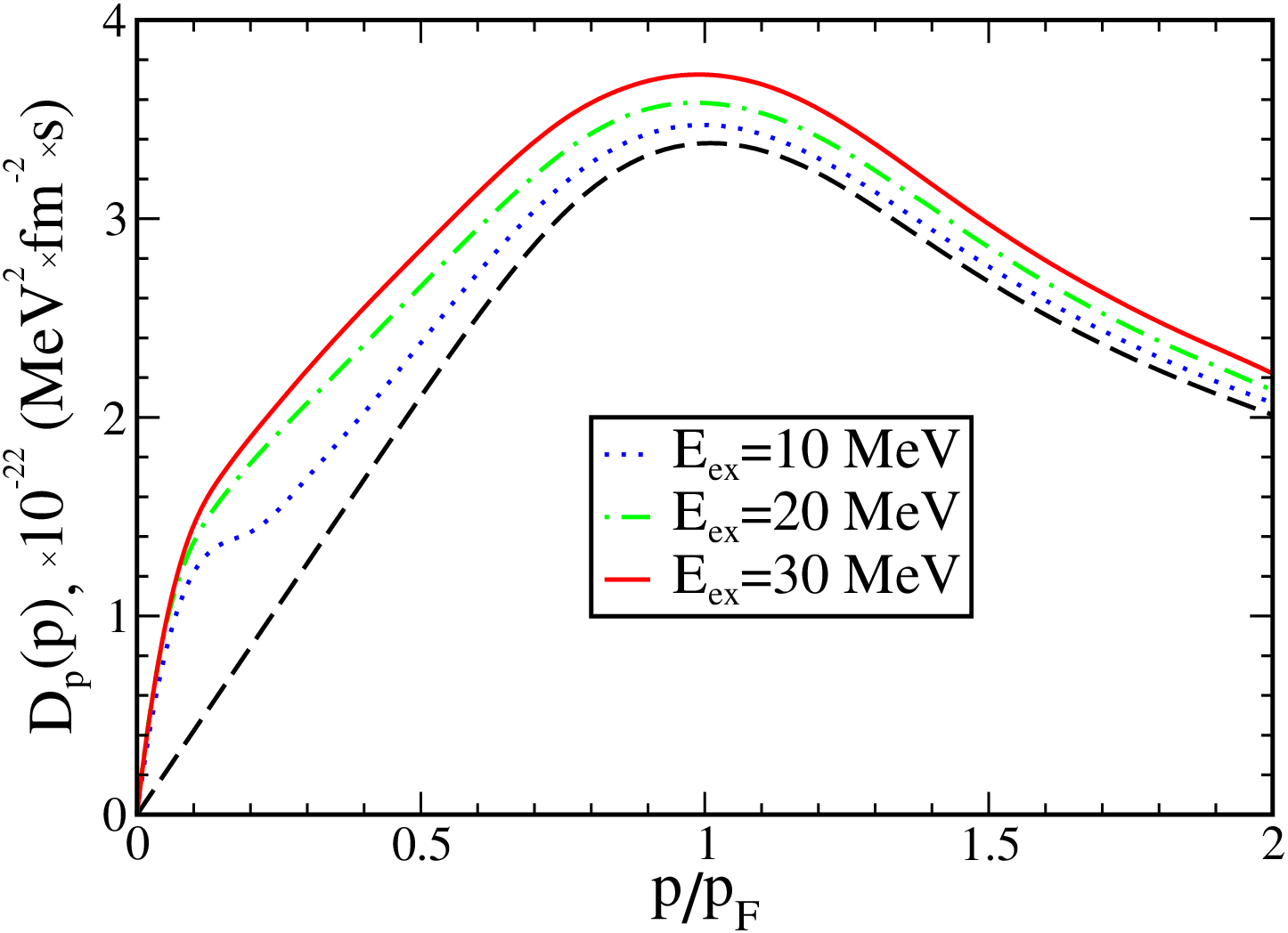}
\caption{Dependence of the diffusion coefficient $D_p(p)$ on the relative momentum $p/p_F$ in the case of excitation of a particle-hole pair 
for a Fermi system which is described by the distribution function (\ref{fus}) for case (b).}
\label{fig5}
\end{center}
\end{figure}

\begin{figure}
\begin{center}
\includegraphics[scale=0.5,clip]{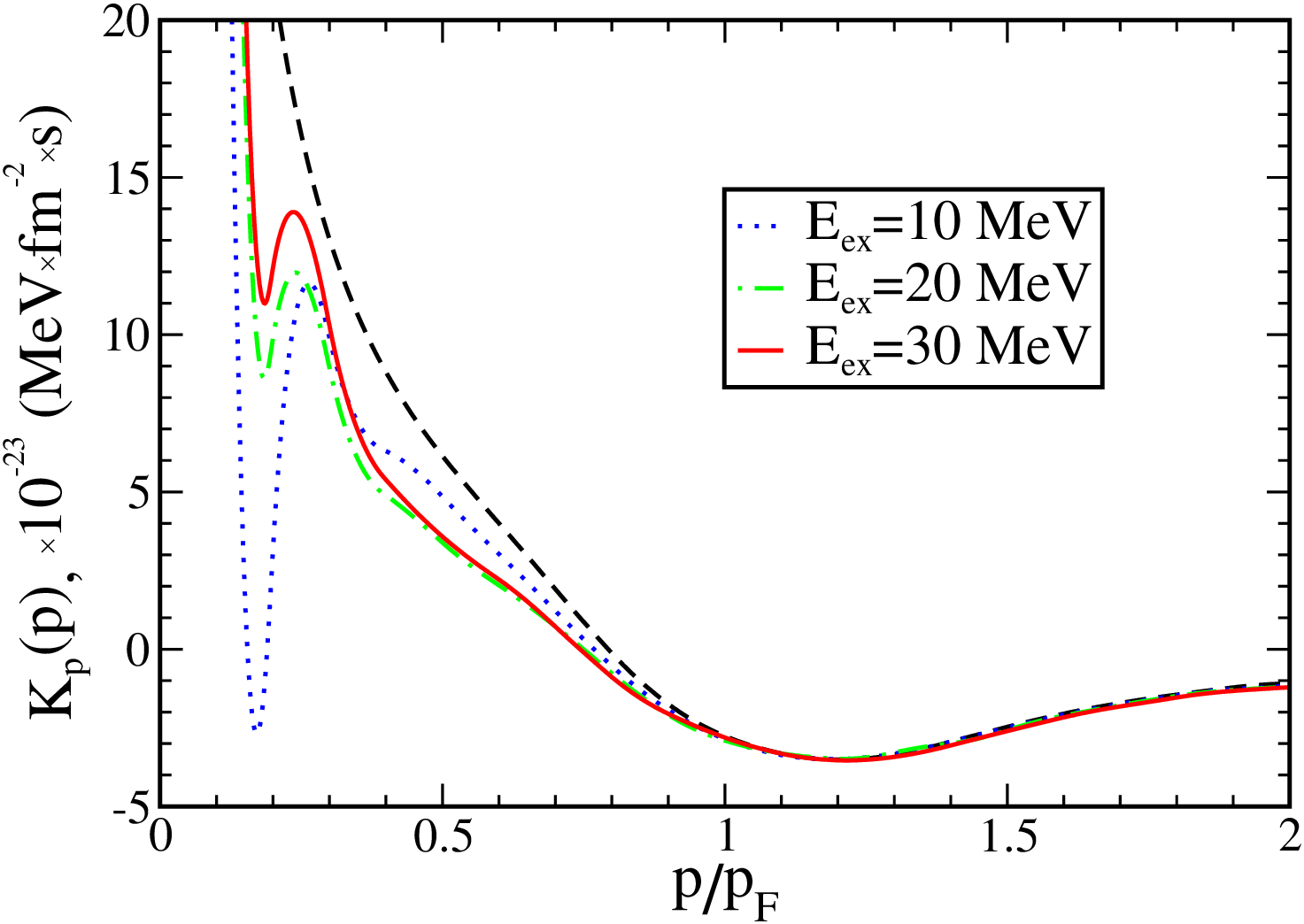}
\caption{Dependence of the drift coefficient $K_p(p)$ on the relative momentum $p/p_F$ in the case of excitation of a particle-hole pair 
for a Fermi system which is described by the distribution function (\ref{fus}) for case (b).}
\label{fig6}
\end{center}
\end{figure}

We have calculated the dependences of the diffusion $D_p(p)$ and drift $K_p(p)$ coefficients on the momentum according to the expressions 
(\ref{dpint0})-(\ref{apik}). Two characteristic cases of excitation of a particle-hole pair are considered for three values of the excitation energy:
$E_{ex}=10, 20$ and $30$ MeV. In the first case (a), the nucleon is excited from a level lying below the Fermi energy by 5 MeV to different levels 
lying above the Fermi energy. In the second case (b), the nucleon is excited from different levels below the Fermi level to the same level 
which lies above the Fermi level by 5 MeV. \figurename~\ref{fig3} and \ref{fig4} show the results of calculations for case (a), 
and \figurename~\ref{fig5} and \ref{fig6} show calculations for case (b).

In general, as can be seen from \figurename~\ref{fig3}, the dependence on the momentum has a similar character to the case of the step distribution 
function which is added here for comparison as a dashed curve. It is also noteworthy that the higher the value of the excitation energy $E_{ex}$ 
the greater the value of $D_p(p)$ except for a small region of small impulses.

In \figurename~\ref{fig4} we present the dependences obtained for the drift coefficient $K_p(p)$ together with the case for the stepwise distribution. 
As can be seen from the figure, $K_p(p)$ is actually independent of the excitation energy outside the Fermi surface (all curves coincide).
In the middle of the Fermi sphere (for $p/p_F<1$) there is a noticeable strong oscillating deviation from the case of the stepwise distribution.
Here one should note the appearance of one more additional minimum at values of the relative momentum of approximately $p/p_F\approx 0.18$.
It is interesting to note that the position of this minimum does not depend on the excitation energy but its depth decreases 
(the absolute value increases) with increasing $E_{ex}$.
From a comparison of the distribution (\ref{fin}) and the stepwise distribution (\ref{fus}), we conclude that the deviation in the dependencies
in \figurename~\ref{fig4} from the dashed line and the appearance of irregular oscillations is obviously a manifestation of particle-hole excitation
in the Fermi system.

The calculation results for the second case (b) are shown in \figurename~\ref{fig5} and \figurename~\ref{fig6}.
As we can see, the behavior of both kinetic coefficients does not qualitatively differ from the first case (a).
All comments and conclusions made for case (a) are also valid.
It should only be noted here that the growth of the diffusion coefficient with the excitation energy occurs somewhat more intensively.

\begin{figure}
\begin{center}
\includegraphics[scale=0.5,clip]{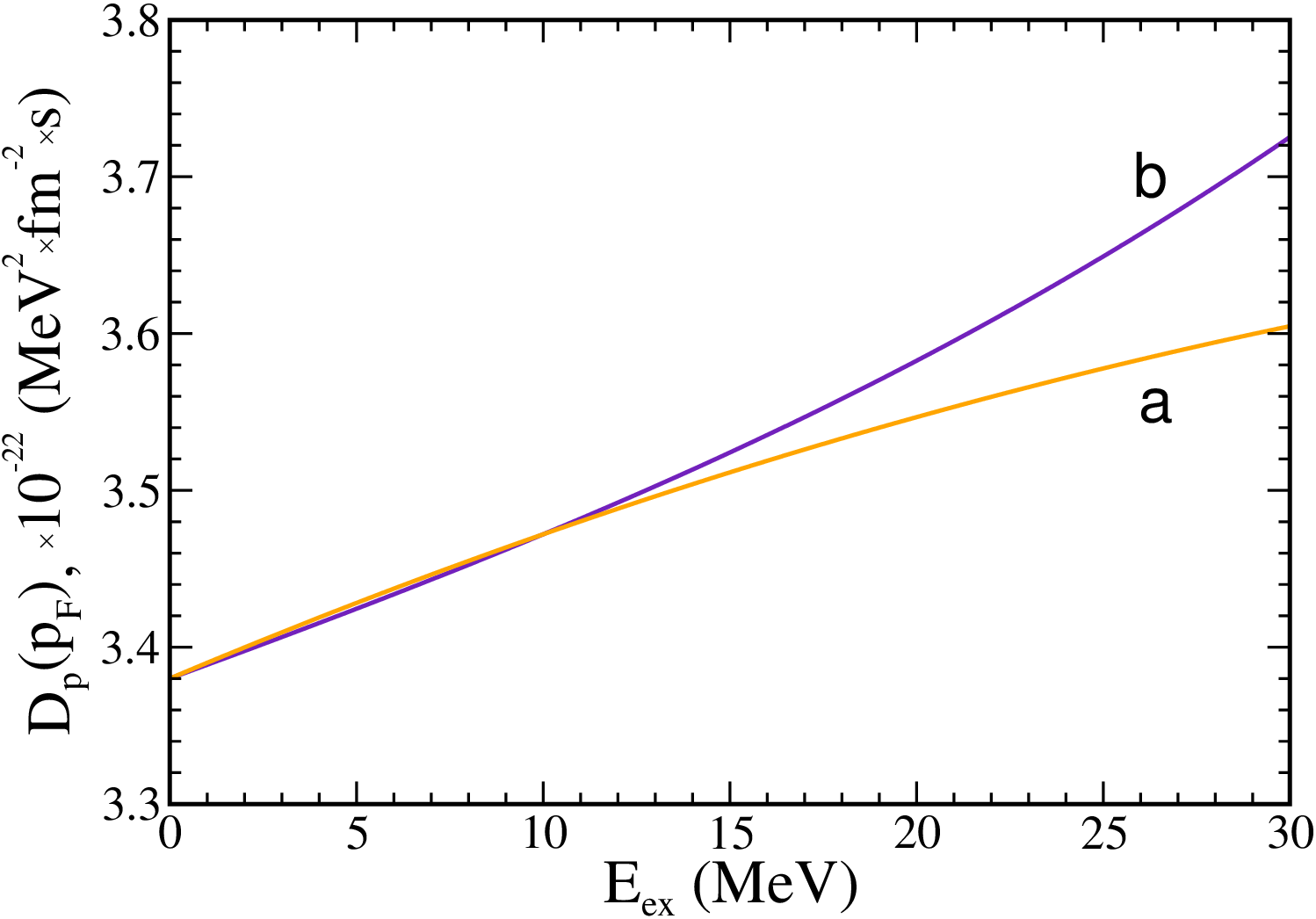}
\caption{Dependence of the diffusion coefficient $D_p(p_F)$ on the excitation energy $E_{ex}$ of a particle-hole pair for a momentum
$p=p_F$. The lower curve is the case (a), the upper curve is (b).}
\label{fig7}
\end{center}
\end{figure}

In \figurename~\ref{fig7} we have shown for both cases the dependence of the maximum value of the diffusion coefficient $D_p(p_F)$ 
located on the Fermi surface on the excitation energy $E_{ex}$.
Case (a) corresponds to the lower curve, case (b) to the upper one. These dependencies can be approximated using the expression
\begin{equation}
D_p(p_F)=D_{p,0}(p_F)\left(1+c_1\cdot E_{ex}+c_2\cdot E_{ex}^2\right),
\label{fit}
\end{equation}
where the fitting coefficients are given in \tablename~\ref{tab1}.
\begin{table}[ht]
\caption{Fitting coefficients.}
\centering
\begin{tabular}{c c c}
\hline\hline
Cases\ \ \ \ &\ \ \ \ $c_1$ [MeV$^{-1}$]\ \ \ \ &\ \ \ \ $c_2$ [MeV$^{-2}$]\ \ \ \ \\ [0.5ex]
\hline 
(a) & $2.97\cdot 10^{-3}$ & $-2.51\cdot 10^{-5}$ \\
(b) & $2.27\cdot 10^{-3}$ & $\ \ 3.75\cdot 10^{-5}$ \\ [1ex]
\hline
\end{tabular}
\label{tab1}
\end{table}
The coefficients $c_1$ for both cases have close values, but the coefficients at quadratic terms $c_2$ have opposite signs.
Thus, the quadratic dependence of the diffusion coefficient on the excitation energy has a directly opposite dependence in both cases.

\section{Conclusions}

In this paper the properties of the kinetic diffusion $D_p(p)$ and drift $K_p(p)$ coefficients in the momentum space for a cold Fermi system 
and also for the case of excitation of a particle-hole pair in this system are considered for the first time. We have used the fact that 
when integrating in the expressions for the kinetic coefficients the stepwise distribution function highlights certain regions in the momentum space, 
that allow rewriting the twofold integral expressions so that instead of the distribution functions, certain limits of integration can be introduced into them.
This gives a significant acceleration in numerical calculations and increases their accuracy.

As a result of numerical calculations, it was found that in the cold Fermi system the diffusion coefficient $D_p(p)$ is different from zero and has a positive
maximum on the Fermi surface, at zero momentum it is zero and at an infinitely large increase in momentum it approaches zero asymptotically.
The drift coefficient $K_p(p)$ is also not zero. It has a negative minimum and approaches zero asymptotically as the relative momentum increases.
When the momentum decreases to zero $K_p(p)$ becomes positive and tends to infinity. The obtained values of the coefficients $D_{p,0}(p_F)$ and $K_{p,0}(p_F)$ 
are in agreement with the values of the corresponding quantities given in the paper \cite{Wo.PRL.1982}.

The difference from zero of the diffusion and drift coefficients means that even in an absolutely cold Fermi system momentum transfer processes 
are possible and consequently change of state. This change will occur as quickly as possible in the region of the Fermi surface and with distance 
from it the rate of smearing of the distribution function will decrease.

In the case of excitation of a particle-hole pair two characteristic cases are considered for three values of the excitation energy.
In general, the dependence of $D_p(p)$ on the momentum is similar to the case of the stepwise distribution. It is also noticeable that the higher 
the energy value excitation $E_{ex}$ the larger $D_p(p)$ except for a small region of small momenta. The kinetic drift coefficient $K_p(p)$ inside 
the Fermi sphere exhibits a strong oscillating deviation from the case of the stepwise distribution and appears another additional minimum. 
The position of this minimum does not depend on the excitation energy but its depth decreases (absolute value increases) with increasing $E_{ex}$.

The obtained dependences of the kinetic diffusion and drift coefficients in momentum space will be useful in further studying the evolution of the distribution function and also the processes of collective motions dissipation using the nonlinear diffusion equation.

\section{Acknowledgments}

Thanks to the Armed Forces of Ukraine for providing security during this work.
This work was supported in part by the budget program ”Support for the development of priority areas of scientific researches”, 
the project of the Academy of Sciences of Ukraine (Code 6541230, no. 0122U000848)

\appendix\section{General transformations of integral expressions}

Given the relationship between energy and momentum $\epsilon_{j}=\mathbf{p}_{j}^{2}/2m$, the argument of the delta function can be rewritten as follows
$$
\epsilon_{2}-\epsilon_{4}-\frac{\mathbf{p}}{m}(\mathbf{p}_{2}-\mathbf{p}_{4})
=\frac{1}{2m}\left((\mathbf{p}_{2}-\mathbf{p})^{2}-(\mathbf{p}_{4}-\mathbf{p})^{2}\right),
$$
where it is taken into account that $\mathbf{p}\equiv\mathbf{p}_{1}$.

We will introduce the new variables $\mathbf{p}_{2}-\mathbf{p}=\mathbf{q}$ and $\mathbf{p}_{4}
-\mathbf{p}=\mathbf{k}$ which allows one to rewrite Eqs. (\ref{dp2}), (\ref{ap2}) in the following form
\begin{equation}
D_p(\mathbf{p})\approx\frac{2g^2}{3m(2\pi\hbar)^{3}} \int d\mathbf{k}\
d\mathbf{q}\ (\mathbf{q}-\mathbf{k})^2\
\frac{d\sigma }{d\Omega } \left((\mathbf{q}-\mathbf{k})^2\right)\tilde{f}(\mathbf{q}+\mathbf{p})f(\mathbf{k}+\mathbf{p}) 
\delta \left(\mathbf{q}^2-\mathbf{k}^2\right),
\label{dp3}
\end{equation}
\begin{equation}
A(\mathbf{p}) \approx \frac{4g^2}{m(2\pi\hbar)^{3}} \int d \mathbf{k}\
d\mathbf{q}\ \hat{\mathbf{p}}(\mathbf{q}-\mathbf{k})\
\frac{d\sigma }{d\Omega } \left((\mathbf{q}-\mathbf{k})^2\right)
\tilde{f}(\mathbf{q}+\mathbf{p})f(\mathbf{k}+\mathbf{p}) 
\delta \left(\mathbf{q}^2-\mathbf{k}^2\right).
\label{ap3}
\end{equation}
Using the spherically symmetric distribution function $f(\mathbf{p})=f(p)$, $\mathbf{p}=p$ and the relation \cite{Dav.b.65}
$$
\delta\left(x^{2}-a^{2}\right)=\frac{\delta(x-a)+\delta(x+a)}{2|a|},
$$
we will rewrite Eqs. (\ref{dp3}), (\ref{ap3}) as
\begin{eqnarray}
D_p(p)&\approx &\frac{g^2}{3m(2\pi\hbar)^{3}} 
\int_{0}^{\infty} kdk\ \int d\Omega _{k} \int_{0}^{\infty} q^{2} dq\ \int d\Omega_{q} 
\nonumber \\
&&\times\left.\left.\left(q^{2}+k^{2}-2qk\cos\widehat{\mathbf{q}\mathbf{k}}\right) 
\frac{d\sigma}{d\Omega}\left(q^{2}+k^{2}-2qk\cos\widehat{\mathbf{q}\mathbf{k}}\right)
\right\{\delta(q-k)+\delta(q+k)\right\}
\nonumber \\
&&\times\tilde{f}\left(\sqrt{q^{2}+p^{2}+2qp\cos\widehat{\mathbf{q}\mathbf{p}}}\right) 
f\left(\sqrt{k^{2}+p^{2}+2kp\cos\widehat{\mathbf{k}\mathbf{p}}}\right),
\label{dp4}
\end{eqnarray}
\begin{eqnarray}
A(p)&\approx &\frac{2g^2}{m(2\pi\hbar)^{3}} 
\int_{0}^{\infty} kdk\ \int d\Omega _{k}
\int_{0}^{\infty} q^{2} dq\ \int d\Omega_{q} 
\nonumber \\
&&\times\left.\left.\left(q\cos\widehat{\mathbf{q}\mathbf{p}}-k\cos\widehat{\mathbf{k}\mathbf{p}}\right) 
\frac{d\sigma}{d\Omega}\left(q^{2}+k^{2}-2qk\cos\widehat{\mathbf{q}\mathbf{k}}\right)
\right\{\delta(q-k)+\delta(q+k)\right\}
\nonumber \\
&&\times \tilde{f}\left( \sqrt{q^{2}+p^{2}+2qp\cos\widehat{\mathbf{q}\mathbf{p}}}\right) 
f\left(\sqrt{k^{2}+p^{2}+2kp\cos\widehat{\mathbf{k}\mathbf{p}}}\right).
\label{ap4}
\end{eqnarray}

Integrating in Eqs. (\ref{dp4}), (\ref{ap4}) over $q$, we obtain
\begin{eqnarray}
D_p(p)&\approx&\frac{2g^2}{3m(2\pi\hbar)^{3}} 
\int_{0}^{\infty} k^5 dk\ \int d\Omega_{k}\ 
f\left(\sqrt{k^2+p^2+2kp\cos\widehat{\mathbf{k}\mathbf{p}}}\right)
\nonumber \\
&&\times\int d\Omega_{q}\ 
\left\{\left(1-\cos\widehat{\mathbf{q}\mathbf{k}}\right) 
\frac{d\sigma}{d\Omega}\left(2k^2(1-\cos\widehat{\mathbf{q}\mathbf{k}})\right)\right.
\tilde{f}\left(\sqrt{k^2+p^2+2kp\cos\widehat{\mathbf{q}\mathbf{p}}}\right)
\nonumber \\
&&+\left(1+\cos\widehat{\mathbf{q}\mathbf{k}}\right) 
\frac{d\sigma}{d\Omega}\left(2k^2(1+\cos\widehat{\mathbf{q}\mathbf{k}})\right)
\left.\tilde{f}\left( \sqrt{k^{2}+p^{2}-2kp
\cos\widehat{\mathbf{q}\mathbf{p}}}\right) \right\}
\label{dp5}
\end{eqnarray}
\begin{eqnarray}
A(p)&\approx&\frac{2g^2}{m(2\pi\hbar)^{3}} 
\int_{0}^{\infty} k^4 dk\ \int d\Omega_{k}\ 
f\left(\sqrt{k^2+p^2+2kp\cos\widehat{\mathbf{k}\mathbf{p}}}\right)
\nonumber \\
&&\times \int d\Omega_{q}\ \left\{
\left(\cos\widehat{\mathbf{q}\mathbf{p}}-\cos\widehat{\mathbf{k}\mathbf{p}}\right) 
\frac{d\sigma}{d\Omega}\left(2k^2(1-\cos\widehat{\mathbf{q}\mathbf{k}})\right)\right.
\tilde{f}\left(\sqrt{k^2+p^2+2kp\cos\widehat{\mathbf{q}\mathbf{p}}}\right)
\nonumber \\
&&-\left(\cos\widehat{\mathbf{q}\mathbf{p}}+\cos\widehat{\mathbf{k}\mathbf{p}}\right) 
\frac{d\sigma}{d\Omega}\left(2k^2(1+\cos\widehat{\mathbf{q}\mathbf{k}})\right)
\left.\tilde{f}\left(\sqrt{k^{2}+p^{2}-2kp\cos\widehat{\mathbf{q}\mathbf{p}}}\right) \right\}.
\label{ap5}
\end{eqnarray}

Let us choose the direction of the $z$ axis of the coordinate system in the direction of the vector $\mathbf{p}$.
Using the addition theorem for spherical harmonics, we can write in an arbitrary spherical coordinate system
\begin{eqnarray}
\cos\widehat{\mathbf{q}\mathbf{k}}=
\cos\theta_{q}\cos\theta_{k}+\sin\theta_{q}\sin\theta_{k}\cos(\phi_{q}+\phi_{k}),  
\nonumber
\\
\cos\widehat{\mathbf{q}\mathbf{p}}=
\cos\theta_{q}\cos\theta_{p}+\sin\theta_{q}\sin\theta_{p}\cos(\phi_{q}+\phi_{p}),
\nonumber \\
\cos\widehat{\mathbf{k}\mathbf{p}}=
\cos\theta_{k}\cos\theta_{p}+\sin\theta_{k}\sin\theta_{p}\cos(\phi_{k}+\phi_{p}).
\nonumber
\end{eqnarray}
Taking into account the spherically symmetry of the distribution functions $f(\mathbf{p}_{j})=f(p_{j})$ and using
$$
\cos\widehat{\mathbf{q}\mathbf{p}}=\cos\theta_q,
\qquad 
\cos\widehat{\mathbf{k}\mathbf{p}}=\cos\theta_k,
$$
we finally obtain
\begin{eqnarray}
D_p(p)&\approx&\frac{2g^2}{3m(2\pi\hbar)^{3}} 
\int_{0}^{\infty} k^5 dk\ \int d\Omega_{k}\ f\left(\sqrt{k^2+p^2+2kp\cos\theta_k}\right)
\nonumber \\
&& \times \int d\Omega_{q} 
\left\{\left(1-\cos\widehat{\mathbf{q}\mathbf{k}}\right) 
\frac{d\sigma}{d\Omega}\left(2k^2(1-\cos \widehat{\mathbf{q}\mathbf{k}})\right)\right.
\tilde{f}\left(\sqrt{k^2+p^2+2kp\cos\theta_q}\right)
\nonumber \\
&& + \left(1+\cos\widehat{\mathbf{q}\mathbf{k}}\right) 
\frac{d\sigma}{d\Omega}\left(2k^2(1+\cos\widehat{\mathbf{q}\mathbf{k}})\right)
\left.\tilde{f}\left(\sqrt{k^{2}+p^{2}-2kp\cos\theta_q}\right)\right\},
\label{dp6}
\end{eqnarray}
\begin{eqnarray}
A(p)&\approx&\frac{2g^2}{m(2\pi\hbar)^{3}} 
\int_{0}^{\infty} k^4 dk\ \int d\Omega_{k}\ f\left(\sqrt{k^2+p^2+2kp\cos\theta_k}\right)
\nonumber \\
&& \times \int d\Omega _{q}\ \left\{\left(\cos\theta_q-\cos\theta_k\right) 
\frac{d\sigma}{d\Omega}\left(2k^2(1-\cos \widehat{\mathbf{q}\mathbf{k}})\right)\right.
\tilde{f}\left(\sqrt{k^2+p^2+2kp\cos\widehat{\mathbf{q}\mathbf{p}}}\right)
\nonumber \\
&& -\left(\cos\theta_q+\cos\theta_k\right) 
\frac{d\sigma}{d\Omega}\left(2k^2(1+\cos \widehat{\mathbf{q}\mathbf{k}})\right)
\left.\tilde{f}\left(\sqrt{k^{2}+p^{2}-2kp\cos\widehat{\mathbf{q}\mathbf{p}}}\right) 
\right\}.
\label{a6}
\end{eqnarray}

\section{Transformation of integral expressions in the case of the Gauss potential}

After substituting (\ref{pot}) into the integral expressions (\ref{dp6-a}) and (\ref{a6-a}), we get
\begin{eqnarray}
D_p(p)&\approx&\frac{g^2 m r_{0}^{6}v_{0}^{2}}{24\pi^2\hbar^7}
\int_{0}^{\infty} k^5 dk \int d\Omega_{k}\ f\left(\sqrt{k^2+p^2+2kp\cos\theta_k}\right)
\nonumber \\
&& \times\int d\Omega_{q}
\left\{\left(1-\cos\widehat{\mathbf{q}\mathbf{k}}\right) 
\exp\left(-8k^2r_0^2(1-\cos \widehat{\mathbf{q}\mathbf{k}})/\hbar^2\right)
\tilde{f}\left(\sqrt{k^2+p^2+2kp\cos\theta_q}\right)\right.
\nonumber \\
&& +\left.\left(1+\cos \widehat{\mathbf{q}\mathbf{k}}\right) 
\exp\left(-8k^2r_0^2(1+\cos\widehat{\mathbf{q}\mathbf{k}})/\hbar^2\right)
\tilde{f}\left( \sqrt{k^{2}+p^{2}-2kp\cos\theta_q}\right) 
\right\},
\label{dp7}
\end{eqnarray}
\begin{eqnarray}
A(p)&\approx&\frac{g^2mr_{0}^{6}v_{0}^{2}}{8\pi^2\hbar^7}
\int_{0}^{\infty} k^4 dk \int d\Omega_{k}\ f\left(\sqrt{k^2+p^2+2kp\cos\theta_k}\right)
\nonumber \\
&& \times\int d\Omega_{q} \left\{\left(\cos\theta_q-\cos\theta_k\right)  
\exp\left(-8k^2r_0^2(1-\cos \widehat{\mathbf{q}\mathbf{k}})/\hbar^2\right)
\tilde{f}\left(\sqrt{k^2+p^2+2kp\cos\theta_q}\right)\right.
\nonumber \\
&& -\left.\left(\cos\theta_q+\cos\theta_k\right) 
\exp\left(-8k^2r_0^2(1+\cos \widehat{\mathbf{q}\mathbf{k}})/\hbar^2\right)
\tilde{f}\left(\sqrt{k^{2}+p^{2}-2kp\cos\theta_q}\right) \right\}.
\label{ap7}
\end{eqnarray}

We introduce the notation
$$
\cos\theta_q=x, \quad \cos\theta_k=y,
$$
then
\begin{eqnarray}
D_p(p)&\approx&\frac{g^2 m r_{0}^{6}v_{0}^{2}}{24\pi^2\hbar^7}
\int_{0}^{\infty} k^5 dk \int_{-1}^1dx \int_{-1}^1dy\ f\left(\sqrt{k^2+p^2+2kpy}\right)
\nonumber \\
&& \times \int_0^{2\pi} d\phi_{k}\ \int_0^{2\pi} d\phi_{q}\
\left\{\left(1-xy-\sqrt{(1-x^2)(1-y^2)}\cos(\phi_k+\phi_q)\right)\right.
\nonumber \\
&&\times\exp\left(-8k^2r_0^2\left(1-xy-\sqrt{(1-x^2)(1-y^2)}\cos(\phi_k+\phi_q)\right)/\hbar^2\right)
\nonumber \\
&&\times \tilde{f}\left(\sqrt{k^2+p^2+2kpx}\right)
+ \left(1+xy+\sqrt{(1-x^2)(1-y^2)}\cos(\phi_k+\phi_q)\right)
\nonumber \\
&& \times\exp\left(-8k^2r_0^2\left(1+xy+\sqrt{(1-x^2)(1-y^2)}\cos(\phi_k+\phi_q)\right)/\hbar^2\right)
\nonumber \\
&& \left.\times\tilde{f}\left( \sqrt{k^{2}+p^{2}-2kpx}\right) 
\right\},
\label{dp8}
\end{eqnarray}
\begin{eqnarray}
A(p)&\approx&\frac{g^2 m r_{0}^{6} v_{0}^{2}}{8\pi^2\hbar^7}
\int_{0}^{\infty} k^4 dk \int_{-1}^1 dx \int_{-1}^1 dy\ f\left(\sqrt{k^2+p^2+2kpy}\right)
\nonumber \\
&& \times \int_0^{2\pi} d\phi_{k}\ \int_0^{2\pi} d\phi_{q}\ 
\left\{(x-y)\tilde{f}\left(\sqrt{k^2+p^2+2kpx}\right) \right.
\nonumber \\
&& \times\exp\left(-8k^2r_0^2\left(1-xy-\sqrt{(1-x^2)(1-y^2)}\cos(\phi_k+\phi_q)\right)/\hbar^2\right)
\nonumber \\
&& -(x+y) \tilde{f}\left(\sqrt{k^{2}+p^{2}-2kpx}\right)
\nonumber \\
&&\times \left.
\exp\left(-8k^2r_0^2\left(1+xy+\sqrt{(1-x^2)(1-y^2)}\cos(\phi_k+\phi_q)\right)/\hbar^2\right) 
\right\}.
\label{ap8}
\end{eqnarray}

Then
\begin{eqnarray}
D_p(p)&\approx&\frac{g^2 m r_{0}^{6}v_{0}^{2}}{24\pi^2\hbar^7}
\int_{0}^{\infty} k^5 dk \int_{-1}^1dx \int_{-1}^1dy\ f\left(\sqrt{k^2+p^2+2kpy}\right)
\nonumber \\
&&\times \left\{
\exp\left(-8k^2r_0^2(1-xy)/\hbar^2\right) \tilde{f}\left(\sqrt{k^2+p^2+2kpx}\right)
\right.
\nonumber \\
&& \times \int_0^{2\pi} d\phi_{k}\ \int_0^{2\pi} d\phi_{q}\
\left(1-xy-\sqrt{(1-x^2)(1-y^2)}\cos(\phi_k+\phi_q)\right)
\nonumber 
\end{eqnarray}
\begin{eqnarray}
&& \times\exp\left(8k^2r_0^2\sqrt{(1-x^2)(1-y^2)}/\hbar^2\cos(\phi_k+\phi_q)\right)
\nonumber \\
&& + \exp\left(-8k^2r_0^2(1+xy)/\hbar^2\right)
\tilde{f}\left(\sqrt{k^2+p^2-2kpx}\right)
\nonumber \\
&& \times \int_0^{2\pi} d\phi_{k}\ \int_0^{2\pi} d\phi_{q}\
\left(1+xy+\sqrt{(1-x^2)(1-y^2)}\cos(\phi_k+\phi_q)\right)
\nonumber \\
&& \times \left.\exp\left(-8k^2r_0^2\sqrt{(1-x^2)(1-y^2)}/\hbar^2\cos(\phi_k+\phi_q)\right)\right\},
\label{dp9}
\end{eqnarray}
\begin{eqnarray}
A(p)&\approx&\frac{g^2 m r_{0}^{6} v_{0}^{2}}{8\pi^2\hbar^7}
\int_{0}^{\infty} k^4 dk \int_{-1}^1 dx \int_{-1}^1 dy\ f\left(\sqrt{k^2+p^2+2kpy}\right)
\nonumber \\
&& \times \left\{ (x-y)\tilde{f}\left(\sqrt{k^2+p^2+2kpx}\right) 
\exp\left(-8k^2r_0^2(1-xy)/\hbar^2\right) \right.
\nonumber \\
&& \times \int_0^{2\pi} d\phi_{k}\ \int_0^{2\pi} d\phi_{q}\ 
\exp\left(8k^2r_0^2\sqrt{(1-x^2)(1-y^2)}/\hbar^2 \cos(\phi_k+\phi_q) \right)
\nonumber \\
&& -(x+y)\tilde{f}\left(\sqrt{k^2+p^2-2kpx}\right)
\exp\left(-8k^2r_0^2(1+xy)/\hbar^2\right)
\nonumber \\
&& \times \left. \int_0^{2\pi} d\phi_{k}\ \int_0^{2\pi} d\phi_{q}\ 
\exp\left(-8k^2r_0^2\sqrt{(1-x^2)(1-y^2)}/\hbar^2\cos(\phi_k+\phi_q)\right)
\right\}.
\label{ap9}
\end{eqnarray}

To simplify, we introduce the following notation
\begin{equation}
q_{\pm x}^2=p^2+k^2\pm 2pkx, \quad q_{\pm y}^2=p^2+k^2\pm 2pky.
\end{equation}
\begin{equation}
\alpha_{\pm}(k,x,y)=8k^2r_0^2(1\pm xy)/\hbar^2, 
\end{equation}
\begin{equation}
\beta(k,x,y)=8k^2r_0^2\sqrt{(1-x^2)(1-y^2)}/\hbar^2.
\end{equation}

Using the definition of modified Bessel functions of the first kind, we obtain for the integrals
$$
\int_{0}^{2\pi}d\phi_{q}\int_{0}^{2\pi }d\phi_{k}\ 
\exp\left(\beta(k,x,y)\cos(\phi_{q}+\phi_{k})\right)=4\pi^2 I_0(\beta(k,x,y)),
$$
$$
\int_{0}^{2\pi}d\phi_{q}\int_{0}^{2\pi}d\phi_{k}\
\cos(\phi_{q}+\phi_{k})\exp\left(\beta(k,x,y)\cos(\phi_{q}+\phi_{k})\right)=4\pi^2 I_1(\beta(k,x,y)).
$$

Then we will finally have
\begin{eqnarray}
&&D_p(p)\approx \frac{g^2 m r_{0}^{6}v_{0}^{2}}{6\hbar^7}
\int_{0}^{\infty} k^5 dk \int_{-1}^1 dx \int_{-1}^1 dy\ f\left(q_{+y}\right)
\nonumber \\
&&\times \left\{\exp(-\alpha_{-}(k,x,y))\tilde{f}(q_{+x}) 
\left[(1-xy)I_0(\beta(k,x,y))-\sqrt{(1-x^2)(1-y^2)}I_1(\beta(k,x,y))\right]\right.
\nonumber \\
&&+\left.\exp(-\alpha_{+}(k,x,y))\tilde{f}(q_{-x})
\left[(1+xy)I_0(-\beta(k,x,y))+\sqrt{(1-x^2)(1-y^2)}I_1(-\beta(k,x,y))\right]\right\},
\nonumber \\
\label{dp11}
\end{eqnarray}
\begin{eqnarray}
A(p)&\approx &\frac{g^2 m r_{0}^{6} v_{0}^{2}}{2\hbar^7}
\int_{0}^{\infty} k^4 dk \int_{-1}^1 dx \int_{-1}^1 dy\ f\left(q_{+y}\right)
\nonumber \\
&&\times \left\{ (x-y) \tilde{f}\left(q_{+x}\right) 
\exp\left(-\alpha_{-}(k,x,y)\right) I_0\left(\beta(k,x,y)\right)\right.
\nonumber \\
&&-\left.(x+y) \tilde{f}\left(q_{-x}\right)
\exp\left(-\alpha_{+}(k,x,y)\right) I_0\left(-\beta(k,x,y)\right)
\right\}. 
\label{ap11}
\end{eqnarray}

Taking into account the symmetry properties of the modified Bessel functions of the first kind
$$
I_0(-\beta)=I_0(\beta), \quad I_1(-\beta)=-I_1(\beta),
$$
we will have
\begin{eqnarray}
&&D_p(p)\approx \frac{g^2 m r_{0}^{6}v_{0}^{2}}{6\hbar^7}
\int_{0}^{\infty} k^5 dk \int_{-1}^1 dx \int_{-1}^1 dy\ f\left(q_{+y}\right)
\nonumber \\
&&\times \left\{\exp(-\alpha_{-}(k,x,y))\tilde{f}(q_{+x}) 
\left[(1-xy)I_0(\beta(k,x,y))-\sqrt{(1-x^2)(1-y^2)}I_1(\beta(k,x,y))\right]\right.
\nonumber \\
&&+\left.\exp(-\alpha_{+}(k,x,y))\tilde{f}(q_{-x})
\left[(1+xy)I_0(\beta(k,x,y))-\sqrt{(1-x^2)(1-y^2)}I_1(\beta(k,x,y))\right]\right\},
\nonumber \\
\label{dp12}
\end{eqnarray}
\begin{eqnarray}
A(p)&\approx &\frac{g^2 m r_{0}^{6} v_{0}^{2}}{2\hbar^7}
\int_{0}^{\infty} k^4 dk \int_{-1}^1 dx \int_{-1}^1 dy\ f\left(q_{+y}\right) I_0\left(\beta(k,x,y)\right)
\nonumber \\
&&\times \left\{ (x-y) \tilde{f}\left(q_{+x}\right) 
\exp\left(-\alpha_{-}(k,x,y)\right)
- (x+y) \tilde{f}\left(q_{-x}\right)
\exp\left(-\alpha_{+}(k,x,y)\right) \right\}. 
\label{ap12}
\end{eqnarray}
When integrating over the variable $x$ in the second term in curly braces we make the replacement $x\rightarrow-x$. Then, taking into account 
the symmetries $\alpha_{+}(k,-x,y)=\alpha_{-}(k,x,y)$ and $\beta(k,-x,y)=\beta(k, x,y)$, we see that the second term is identical to the first.
Therefore, we finally have
\begin{eqnarray}
D_p(p)&\approx &\frac{g^2 m r_{0}^{6}v_{0}^{2}}{3\hbar^7}
\int_{0}^{\infty} k^5 dk \int_{-1}^1 dx \int_{-1}^1 dy\ \exp(-\alpha(k,x,y)) \tilde{f}(q_{x}) f(q_{y})
\nonumber \\
&&\times 
\left[(1-xy)I_0(\beta(k,x,y))-\sqrt{(1-x^2)(1-y^2)}I_1(\beta(k,x,y))\right],
\label{dp13}
\end{eqnarray}
\begin{eqnarray}
A(p)&\approx &\frac{g^2 m r_{0}^{6} v_{0}^{2}}{\hbar^7}
\int_{0}^{\infty} k^4 dk \int_{-1}^1 dx \int_{-1}^1 dy\  \exp(-\alpha(k,x,y)) \tilde{f}(q_{x}) f(q_{y}) 
\nonumber \\
&& \times (x-y) I_0\left(\beta(k,x,y)\right),
\label{ap13}
\end{eqnarray}
where $\alpha_{-}(k,x,y)\equiv\alpha(k,x,y)$, $q_{+x}\equiv q_{x}$, $q_{+y}\equiv q_{y}$.

\section{Expressions for the minimum and maximum values of integration limits}
\noindent
\begin{eqnarray}
&& x_{\max,1}=x_{\max,2}=x_{\max,3}=1, \qquad \qquad \quad \ x_{\min,1}=x_{\min,2}=x_{\min,3}=-1, \nonumber \\
&& y_{\max,1}=\max(-1,\min(1,z(p_1^{\prime}),z(p_F))), \quad y_{\min,1}=-1, \nonumber \\
&& y_{\max,2}=\max(-1,\min(1,z(p_F))), \qquad \quad \ \ y_{\min,2}=\min(1,\max(-1,z(p_2^{\prime}))), \nonumber \\
&& y_{\max,3}=\max(-1,\min(1,z(p_2))), \qquad \quad \ \ \ y_{\min,3}=\min(1,\max(-1,z(p_1))), \nonumber \\
&& x_{\max,4}=x_{\max,5}=x_{\max,6}=\max(-1,\min(1,z(p_1^{\prime}),z(p_F))), \nonumber \\
&& x_{\min,4}=x_{\min,4}=x_{\min,4}=-1, \nonumber \\
&& y_{\max,4}=\max(-1,\min(1,z(p_1^{\prime}),z(p_F))), \quad y_{\min,4}=-1, \nonumber \\
&& y_{\max,5}=\max(-1,\min(1,z(p_F))), \qquad \quad \ \ y_{\min,5}=\min(1,\max(-1,z(p_2^{\prime}))), \nonumber \\
&& y_{\max,6}=\max(-1,\min(1,z(p_2))), \qquad \quad \ \ \ y_{\min,6}=\min(1,\max(-1,z(p_1))), \nonumber \\
&& x_{\max,7}=x_{\max,8}=x_{\max,9}=\max(-1,\min(1,z(p_F))), \nonumber \\
&& x_{\min,7}=x_{\min,8}=x_{\min,9}=\min(1,\max(-1,z(p_2^{\prime}))), \nonumber \\
&& y_{\max,7}=\max(-1,\min(1,z(p_1^{\prime}),z(p_F))), \quad y_{\min,7}=-1, \nonumber \\
&& y_{\max,8}=\max(-1,\min(1,z(p_F))), \qquad \quad \ \ y_{\min,8}=\min(1,\max(-1,z(p_2^{\prime}))), \nonumber \\
&& y_{\max,9}=\max(-1,\min(1,z(p_2))), \qquad \quad \ \ \ y_{\min,9}=\min(1,\max(-1,z(p_1))), \nonumber \\
&& x_{\max,10}=x_{\max,11}=x_{\max,12}=\max(-1,\min(1,z(p_2))), \nonumber \\
&& x_{\min,10}=x_{\min,11}=x_{\min,12}=\min(1,\max(-1,z(p_1))), \nonumber \\
&& y_{\max,10}=\max(-1,\min(1,z(p_1^{\prime}),z(p_F))), \quad y_{\min,10}=-1, \nonumber \\
&& y_{\max,11}=\max(-1,\min(1,z(p_F))), \qquad \quad \ \ y_{\min,11}=\min(1,\max(-1,z(p_2^{\prime}))), \nonumber \\
&& y_{\max,12}=\max(-1,\min(1,z(p_2))), \qquad \quad \ \ \ y_{\min,12}=\min(1,\max(-1,z(p_1))). \nonumber 
\end{eqnarray}

\end{document}